\documentclass[]{agujournal2019}
\usepackage{url} %this package should fix any errors with URLs in refs.
\usepackage[inline]{trackchanges} %for better track changes. finalnew option will compile document with changes incorporated.
\usepackage{soul}
\draftfalse
\journalname{Geophysical Research Letters}

\begin{document}

%% ------------------------------------------------------------------------ %%
%  Title
%
% (A title should be specific, informative, and brief. Use
% abbreviations only if they are defined in the abstract. Titles that
% start with general keywords then specific terms are optimized in
% searches)
%
%% ------------------------------------------------------------------------ %%

% Example: \title{This is a test title}

\title{Evidence of a Polar Cyclone on Uranus from VLA Observations}

%% ------------------------------------------------------------------------ %%
%
%  AUTHORS AND AFFILIATIONS
%
%% ------------------------------------------------------------------------ %%

% Authors are individuals who have significantly contributed to the
% research and preparation of the article. Group authors are allowed, if
% each author in the group is separately identified in an appendix.)

% List authors by first name or initial followed by last name and
% separated by commas. Use \affil{} to number affiliations, and
% \thanks{} for author notes.
% Additional author notes should be indicated with \thanks{} (for
% example, for current addresses).

% Example: \authors{A. B. Author\affil{1}\thanks{Current address, Antartica}, B. C. Author\affil{2,3}, and D. E.
% Author\affil{3,4}\thanks{Also funded by Monsanto.}}

\authors{Alex Akins\affil{1}, Mark Hofstadter\affil{1}, Bryan Butler\affil{2}, A. James Friedson\affil{1}, 
Edward Molter\affil{3}, Marzia Parisi\affil{1}, Imke de Pater\affil{3}}

\affiliation{1}{Jet Propulsion Laboratory, California Institute of Technology, Pasadena, CA, USA}
\affiliation{2}{National Radio Astronomy Observatory, Socorro, NM, USA}
\affiliation{3}{Department of Earth and Planetary Sciences, University of California, Berkeley, CA, USA}
% \affiliation{4}{Fourth Affiliation}

%(repeat as many times as is necessary)

%% Corresponding Author:
% Corresponding author mailing address and e-mail address:

% (include name and email addresses of the corresponding author.  More
% than one corresponding author is allowed in this LaTeX file and for
% publication; but only one corresponding author is allowed in our
% editorial system.)

% Example: \correspondingauthor{First and Last Name}{email@address.edu}

\correspondingauthor{Alex Akins}{alexander.akins@jpl.nasa.gov}

%% Keypoints, final entry on title page.

%  List up to three key points (at least one is required)
%  Key Points summarize the main points and conclusions of the article
%  Each must be 140 characters or fewer with no special characters or punctuation and must be complete sentences

% Example:
% \begin{keypoints}
% \item	List up to three key points (at least one is required)
% \item	Key Points summarize the main points and conclusions of the article
% \item	Each must be 140 characters or fewer with no special characters or punctuation and must be complete sentences
% \end{keypoints}

\begin{keypoints}
\item VLA observations in 2021 and 2022 reveal a bright, compact spot centered at Uranus' pole at several wavelengths
\item Constraints on horizontal temperature and composition gradients necessary to explain the polar emission structure are derived
\item Inferred patterns in temperature, zonal wind speed and trace gas abundance variations are consistent with the presence of a compact cyclonic vortex
\end{keypoints}

%% ------------------------------------------------------------------------ %%
%
%  ABSTRACT and PLAIN LANGUAGE SUMMARY
%
% A good Abstract will begin with a short description of the problem
% being addressed, briefly describe the new data or analyses, then
% briefly states the main conclusion(s) and how they are supported and
% uncertainties.

% The Plain Language Summary should be written for a broad audience,
% including journalists and the science-interested public, that will not have 
% a background in your field.
%
% A Plain Language Summary is required in GRL, JGR: Planets, JGR: Biogeosciences,
% JGR: Oceans, G-Cubed, Reviews of Geophysics, and JAMES.
% see http://sharingscience.agu.org/creating-plain-language-summary/)
%
%% ------------------------------------------------------------------------ %%

%% \begin{abstract} starts the second page

\begin{abstract}
We present observations of Uranus in northern spring with the VLA from 0.7 cm to 5 cm. These observations reveal details in thermal emission from Uranus' north pole at 10s of bars, including a dark collar near 80$^\circ$N and a bright spot at the polar center. The bright central spot resembles observations of polar emission on Saturn and Neptune at shallower pressures. We constrain the variations in temperature and NH$_3$/H$_2$S abundances which could explain these features. We find that the brightness temperature of the polar spot can be recreated through 5 K temperature gradients and/or 10$\times$ depletion of NH$_3$ or H$_2$S vapor between 10-20 bars, both consistent with the presence of a cyclonic polar vortex. The contrast of the polar spot may have increased since 2015, which would suggest seasonal evolution of Uranus' polar circulation at depth.
\end{abstract}

%% ------------------------------------------------------------------------ %%
%
%  TEXT
%
%% ------------------------------------------------------------------------ %%

%%% Suggested section heads:
% \section{Introduction}
%
% The main text should start with an introduction. Except for short
% manuscripts (such as comments and replies), the text should be divided
% into sections, each with its own heading.

% Headings should be sentence fragments and do not begin with a
% lowercase letter or number. Examples of good headings are:

% \section{Materials and Methods}
% Here is text on Materials and Methods.
%
% \subsection{A descriptive heading about methods}
% More about Methods.
%
% \section{Data} (Or section title might be a descriptive heading about data)
%
% \section{Results} (Or section title might be a descriptive heading about the
% results)
%
% \section{Conclusions}

% Hi, astro-ph-leaks ;) 

\section{Introduction}
Uranus’ obliquity affords unique opportunities for polar observing, which can contribute to our general understanding of polar circulation processes on giant planets. Our current understanding of Uranus' polar circulation (we define the pole as  $>$ 60$^\circ$ latitude) suggests net subsidence as the characteristic feature of vertical motion. This pattern is inferred from emission and reflection contrasts in multi-wavelength observations \cite{dePater1989, dePater1991a, Hofstadter2003, Sromovsky2019, Roman2020}, which suggest depletion of absorbing gases at the poles down to 10s of bars. Zonal wind profiles of polar latitudes, estimated nominally at 1.5 bar, have been most recently updated by \citeA{Karkoschka2015} at southern latitudes based on Voyager 2 images and \citeA{Sromovsky2015} at northern latitudes based on Keck observations. Mean zonal wind speeds decay monotonically from a maximum of ~250 m/s near 60$^\circ$ to near-zero at 90$^\circ$, and there appears to be significant asymmetry between the northern and southern pole zonal wind profiles \cite{Karkoschka1998, Sromovsky2005, Sromovsky2015}. This asymmetry, as well as analyses of Voyager gravity measurements, suggests that zonal wind speeds are vertically confined globally and should decay with altitude in the upper troposphere (e.g. see \citeA{Fletcher2020} for a recent review). As new observations and analysis techniques push limitations of spatial resolution and sensitivity, new features have emerged in images of Uranus that complicate this broader picture. For example, complex polar features have been observed at the CH$_4$ weather level, including a multitude of compact bright features in the northern hemisphere suggestive of small-scale vertical convection acting against net subsidence \cite{Sromovsky2012, Sromovsky2015}.

At many wavelengths, this is the first solstice for which ground-based and near-Earth observatories are sufficiently sensitive to resolve fine details in polar thermal emission. This is certainly true for microwave observations with the Very Large Array (VLA), which underwent an order of magnitude bandwidth upgrade in 2012 that significantly improved the achievable sensitivity. Unlike mid-infrared, near-infrared, and visible observations, which have informed our understanding of Uranus' circulation at and above the H$_2$S and CH$_4$ cloud levels, microwave observations are sensitive to thermal emission from deeper in the troposphere (primarily 10-50 bar at the poles). Post-upgrade VLA observations of Uranus are now sufficiently sensitive to resolve faint zonal banding patterns at lower latitudes at the 10 bar level. Observations of Uranus in 2015 described by \citeA{Molter2021}, specifically the Ku band (2 cm) image, provided evidence of latitudinal structure in tropospheric thermal emission at Uranus' north pole. In their 2 cm image, a faint dark collar and compact polar spot appear at 80$^\circ$N and 90$^\circ$N respectively, and these features were either less pronounced or absent in images obtained at other wavelengths during that campaign. Interestingly, \citeA{GEO1988} and \citeA{Hofstadter1990} also found 80$^\circ$S to also be a local brightness minimum in VLA observations made in the 1980s, although the contrast of the collar in these images was at or beneath the background noise level.

As part of an ongoing effort to analyze over 40 years of VLA observations of Uranus, new observations were obtained in late 2021 and early 2022. These observations have confirmed the presence of polar features at several wavelengths that were apparent in the 2 cm 2015 observations and suggest a brightening of the central polar spot. This striking feature is visually similar to mid-infrared observations of Saturn's South Pole in 2004 described by \citeA{Orton2005} and further observed at both poles by Cassini \cite{Fletcher2018}. A polar vortex was also observed on Neptune's south pole in the near- and mid-IR as well as at radio wavelengths \cite{Luszcz-Cook2010, Fletcher2014, DePater2014, Tollefson2021a}. Since polar vortices appear to be common features on planets \cite{Mitchell2021} and the formation of central polar cyclones on ice giant planets have been predicted from modeling studies \cite{Brueshaber2019}, it seems likely that this feature on Uranus also indicates the presence of a compact polar vortex.

In this letter, we describe recent VLA observations of Uranus from 0.7-5 cm and discuss their implications for atmospheric dynamics at the north pole. In Section \ref{sec:obs}, we discuss the reduction and imaging of the data. In Section \ref{sec:analysis}, we infer constraints on the orders of magnitude of tropospheric temperature and composition variations necessary to create the observed polar structure. Additionally, we use the derived gradients in atmospheric temperature to derive a notional zonal wind profile at 50 bar assuming geostrophic balance, suggesting an strengthening of zonal wind with depth immediately surrounding the polar spot. Finally, we provide additional discussion and concluding comments.

\section{Observations} \label{sec:obs}
Here, we provide an overview of the observations and their reduction/imaging. A manuscript is in preparation that will provide further details regarding the reduction and imaging of these and other Uranus observations, and we refer the interested reader to \citeA{Taylor1999} for more information on radio interferometer datasets and their analysis. A table of relevant information for these observation datasets is included in Supporting Information. We obtained observations of Uranus with the VLA when the sub-observer latitude was near 55$^\circ$N. Observations at K (22 GHz, 1.4 cm), Ka (33 GHz, 0.9 cm) and Q (44 GHz, 0.7 cm) band were taken in October 2021 with the array in B configuration, and observations at C (6 GHz, 5 cm), and Ku (15 GHz, 2 cm) bands were taken in March 2022 with the array in the more extended A configuration. All reduction and imaging of the data was conducted using the CASA package \cite{McMullin2007}. The calibrated interferometer correlations (or visibilities) were flagged for frequency-dependent issues in the wideband spectra, and the visibilities were then averaged within 128MHz-wide sub-bands to reduce data volume. Further flagging was conducted to remove time-dependent issues, and limb-darkened disk starting models were fit to the visibility amplitudes \cite{Butler1999}. To form images, these disk models were subtracted from the observations, the residuals were deconvolved using the multi-scale CLEAN algorithm \cite{Cornwell2008}, and the models were then re-added in image space. Iterative phase self-calibration was applied during the imaging process to correct for atmospheric path fluctuations between calibrator scans, concluding with visibility phase compensation intervals on the order of ten times the visibility integration period (two seconds) \cite{Brogan2018}. Multi-term multifrequency synthesis was used to derive high dynamic range images necessary to observe faint spatial contrasts \cite{Sault1994}. Experiments were also performed in which imaging was conducted using a starting model including previously observed equator-pole emission contrasts on Uranus, and we found that the resulting images were consistent with those using the starting model without this contrast. 

\begin{figure}
\centering
\noindent\includegraphics[width=\textwidth]{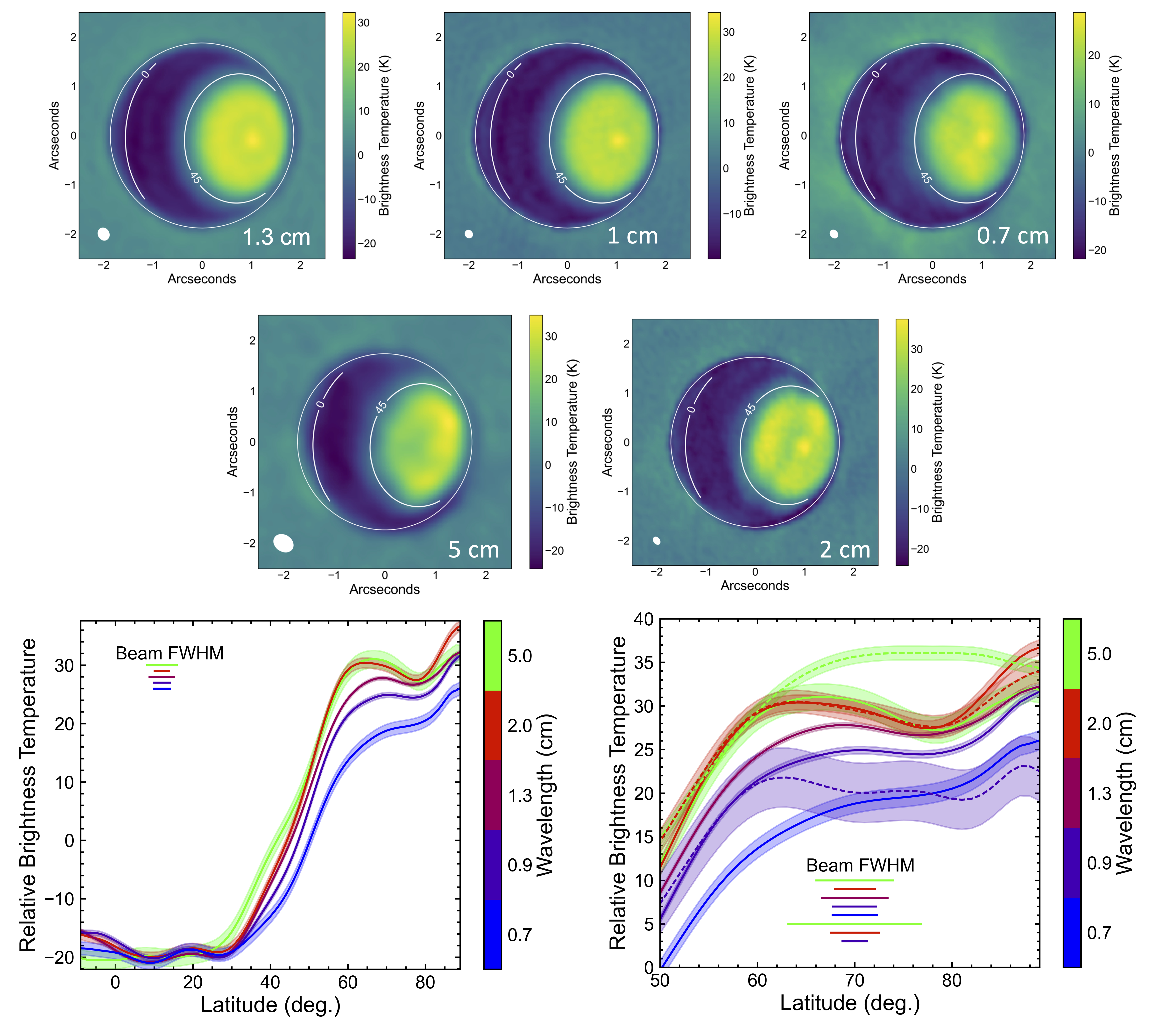}
\caption{(Top) VLA observations of Uranus in October 2021 at K, Ka, and Q band (left to right) and (middle) in March 2022 at C and Ku band (left to right). Details in the spatial brightness temperature structure are illustrated by subtracting a limb-darkened disk model with an equivalent disk-averaged brightness temperature. The image spatial resolution is equivalent to the Gaussian beam shown in the lower left of the images. (Bottom left) Latitudinal trends in disk-subtracted brightness temperature for different observation wavelengths. Profiles are bracketed by 1$\sigma$ uncertainties estimated from the rms noise of the image well outside the disk. The image resolution (synthesized beam full width at half maximum) corresponding to each latitude profile is shown in the top left. (Bottom right) Latitude trends in the polar region for the new observations (solid lines) compared with the results of \citeA{Molter2021} (dashed lines) compensated for differences in emission angle. The image resolution for the 2015 observations are shown below the resolution for the 2022 observations}
\label{fig:uranus_images}
\end{figure}

The final brightness temperature images subtracted by disk models are shown in Figure \ref{fig:uranus_images}. These disk models are weighted by emission angle $\theta$ as $\cos^p\theta$, and the limb-darkening parameter $p$ ranges from $p=0.2$ at 5 cm to $p=0.1$ at 0.7 cm. The sensitivity of the images is at or below 1 K as determined from the root mean square of the emission in the source-free region in all bands, and the absolute calibration of the brightness temperature is on the order 3-5\% \cite{Perley2017}. In addition to the 5 cm image shown in Figure \ref{fig:uranus_images}, another image was made with adjusted visibility weighting to improve the spatial resolution within the diameter of the polar spot at the cost of image sensitivity (see Supporting Information). Latitudinal brightness cross sections for all wavelengths (including the re-weighted high resolution C band image) are also shown at the bottom of Figure \ref{fig:uranus_images}, and these profiles are shifted vertically to obtain alignment between 0-20 degrees latitude. The polar regions of these cross sections are  shown in comparison with the corresponding cross sections of the 2015 observations at C, Ku, and Ka band described by \citeA{Molter2021}. In addition to shifting the cross-sections to match the 2021 and 2022 observations near 60 degrees, the 2015 profiles have also been adjusted to compensate for differences in observation geometry (see Supporting Information). 

The central polar spot is brighter than the 60-70 degree latitude region by 3.5 K in the 2015 2 cm observation, whereas the spot is brighter in the 2021 and 2022 observations by 6-7 K ($>5\sigma$) at several wavelengths. When considering the additional 50\% difference (2.5 degrees latitude) in spatial resolution at 90 degrees between the 2 cm observations in 2015 and 2022, the change in the central spot contrast between the observations is significant on the order of $2\sigma$.  For the re-weighted 5 cm image, the center of the pole does not appear $>1\sigma$ brighter than average of the polar cap. This does not rule out the presence of a brightness contrast at longer wavelengths, only that such a contrast falls below the detection threshold of these observations. The dark trough between 75-80 degrees latitude observed in the 2015 2 cm image is also present in several bands in the new observations, and the collar contrast appears to increase with wavelength. While unable to resolve a polar bright spot, observations with the VLA during the previous south polar solstice (near 1985) also suggest the presence of the dark collar at the limit of the observation sensitivity.  

\section{Analysis} \label{sec:analysis}
To interpret the observations, we use a microwave radiative transfer code and equilibrium cloud-condensation model (ECCM) similar to those employed by \citeA{dePater1985, Hofstadter2003} (see also the RadioBEAR code used by e.g. \citeA{Molter2021, Tollefson2021a}). We assume atmospheric structure derived from the Voyager 2 equatorial radio occultation profile where the temperature is extrapolated using a wet adiabatic lapse rate. At lower latitudes, fit parameters include the deep abundances of trace gases including H$_2$S, NH$_3$, and PH$_3$, which are thought to be the species most likely to affect microwave propagation. Additionally, a relative humidity parameter is fit for condensation of the dominant trace gas (either H$_2$S or NH$_3$) above the NH$_4$SH cloud base located between 15-20 bars.

\begin{figure}
\centering
\noindent\includegraphics[width=\textwidth]{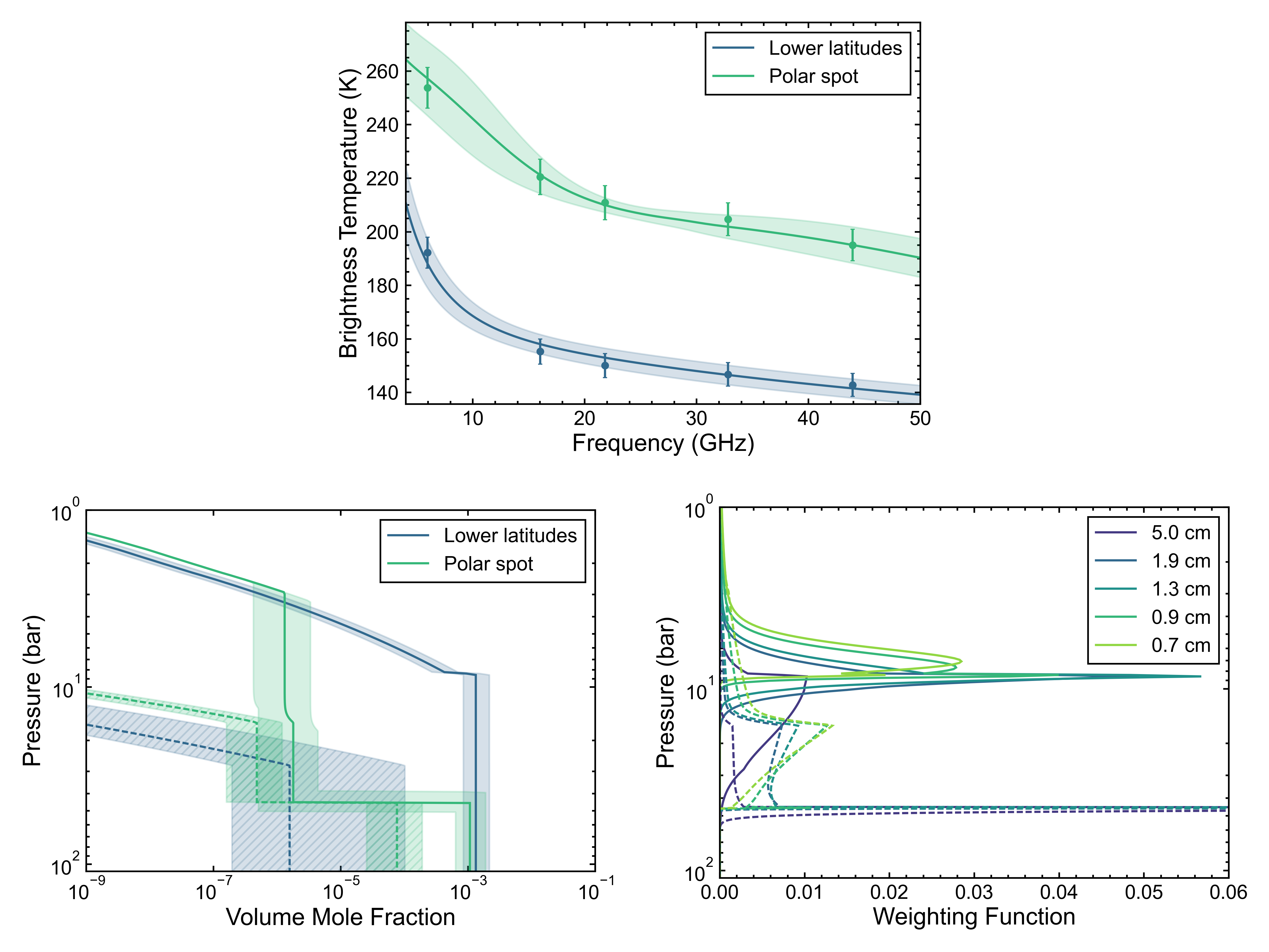}
 \caption{(Top) Brightness temperature spectra of Uranus at lower latitudes ($0^\circ \leq |\theta| \leq 30^\circ$) and the polar spot ($\theta = 90^\circ$) determined from observations. These are compared with the best fit model atmosphere spectra determined via MCMC fit to ECCM parameters with corresponding uncertainties.
 (Bottom left) Best fit lower latitude and polar spot abundances of H$_2$S (solid line) and NH$_3$ (dashed line) and their uncertainties. (Bottom right) Temperature weighting functions for the lower latitude (solid lines) and polar spot (dashed line) fits, demonstrating sensitivity to emission between approximately 5-50 bar.}
\label{fig:vert_prof}
\end{figure}

We are interested in the latitude dependence of the brightness temperature structure, so the observations are binned in latitude and assigned an effective emission angle by averaging over the cosine of the emission angle for each pixel. We perform fits using the ECCM over the lower latitude region ($0^\circ \leq |\theta| \leq 30^\circ$) and multiple polar latitudes using a Markov Chain Monte Carlo approach similar to that discussed in \citeA{Molter2021}. Figure \ref{fig:vert_prof} shows the results of the retrievals over the lower latitudes and at the central polar spot (see Supporting Information for MCMC retrievals at other  latitudes). We find in the equatorial region an H$_2$S-dominant solution with a sub-NH$_4$SH cloud abundance of $1350^{+995}_{-371}$ ppm and H$_2$S relative humidity of $34^{+22}_{-17}$\%. We additionally estimate an upper limit to PH$_3$ abundance at 1.2 ppm. Overall, our fit to the lower-latitude brightness temperature spectrum is consistent within uncertainties with the results of \citeA{Molter2021} for the 2015 observations.

For the polar spot, the nominal ECCM solution is unable match the observed microwave brightness temperature spectrum, so we impose a step boundary layer in the deep atmosphere, following \citeA{Molter2021}. The trace gas abundances below the step are constrained to fall within the range retrieved over the lower latitudes. The inclusion of this step improved the fits to the polar spot spectrum. The pressure at which this step occurred is determined as $45^{+9}_{-7}$ bars, which is consistent with uncertainties with the step location from \citeA{Molter2021}. Due to the Ku-Q band spectral slope, polar ECCM fits require both H$_2$S and NH$_3$ to be present at altitudes above the step with H$_2$S being more abundant. This is in contrast with the retrievals of \citeA{Molter2021} which prefer an NH$_3$ dominant fit in the polar region. We refer the reader to Supporting Information for numerical results of MCMC fits over the polar region and a comparison of fits with and without a deep atmosphere step boundary. 

 In general, the microwave brightness temperature of Uranus depends on the atmospheric temperature and wavelength-dependent opacity of atmospheric gases (see e.g. Figure 8 in \citeA{Molter2021}). Thus, for a given set of measurements, non-unique solutions exist for the vertical temperature structure and vertical distribution of microwave absorbers, and the use of an ECCM constrains the range of achievable solutions. The difficulty in joint absolute determination of atmospheric temperature and composition profiles in the absence of model assumptions from these observations is also a result of the uncertainty in the VLA flux calibration. Whereas statistical uncertainty in the retrieved vertical structure is determined by the flux calibration accuracy, statistical uncertainties in the variations of atmospheric state with latitude are instead determined by the noise in the images, which is approximately ten times smaller. We can therefore determine relatively strong constraints on how the atmospheric state varies with latitude along isobars, which we report here. We consider two end-member cases to explain these observations: 1) the brightness temperature variations are solely caused by changes in atmospheric temperature, or 2) they are solely caused by changes in atmospheric composition, specifically the abundances of NH$_3$ and H$_2$S vapors. For our retrievals, we adjust the profiles of either temperature or gas abundance to match latitude-dependent brightness temperature spectra by scaling the reference polar spot profile at specific points on a coarse grid (see Supporting Information for a discussion of the choice of grid points). 
 
\begin{figure}[htbp]
\begin{center}
\noindent\includegraphics[width=\textwidth]{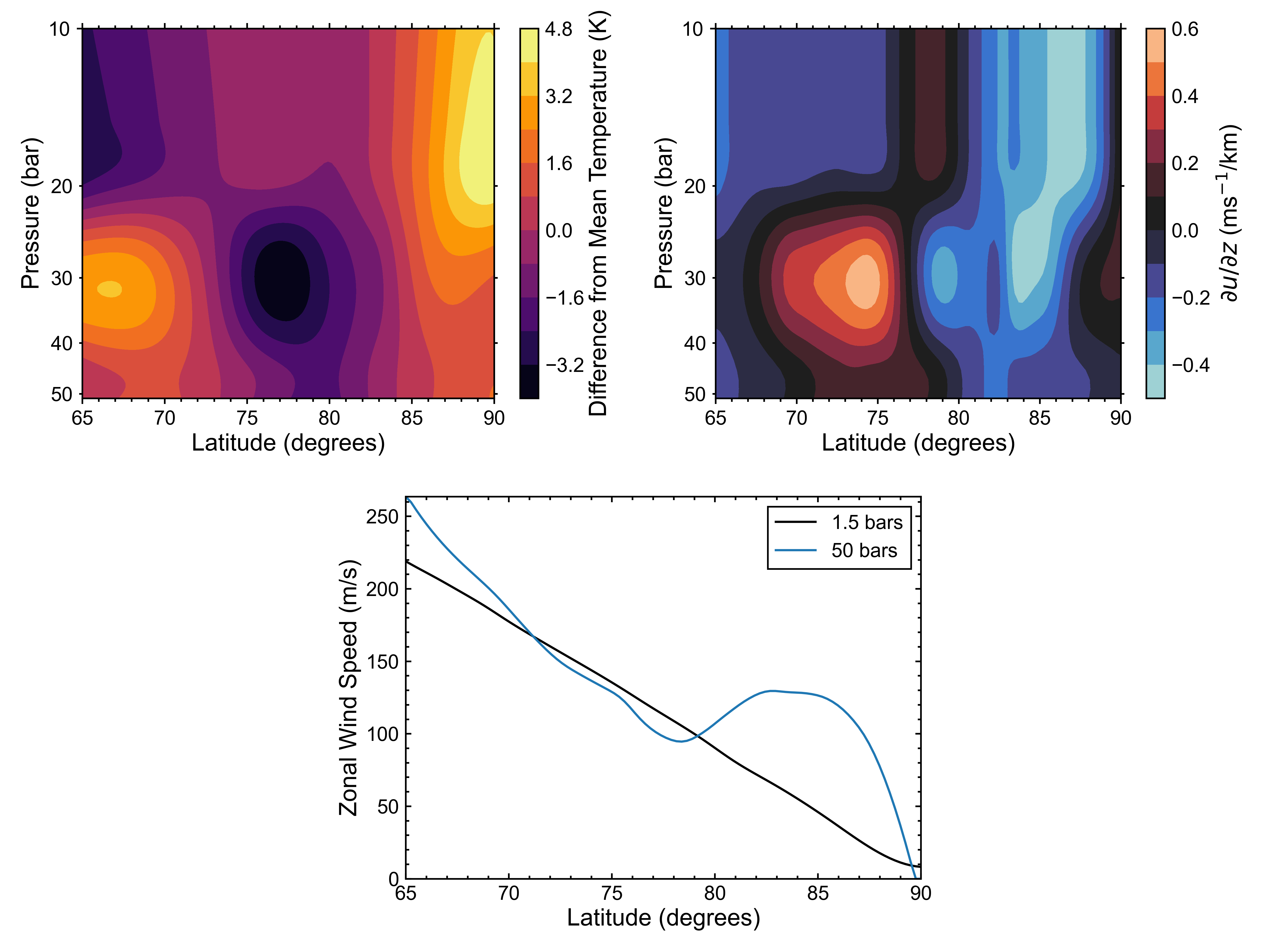}
\caption{(Top left) Deviation of the retrieved polar temperature profiles from the mean (averaged over all latitudes) retrieved profile corresponding to end-model 1 (see text). (Top right) Vertical shear in zonal winds derived using the retrieved temperature variations and the thermal wind equation. For both of the top charts, the spatial resolution of the inferred quantities range between 4-8 degrees of latitude, and the NH$_4$SH condensation level occurs near 18 bar. (Bottom) The zonal wind profile derived by \citeA{Sromovsky2015} at approximately 1.5 bar is integrated over the wind shear profile to approximate the zonal wind speed at 50 bars assuming all variations in polar brightness temperature are the result of changes in atmospheric temperature.}
\label{fig:temp_analysis}
\end{center}
\end{figure}

\begin{figure}[htbp]
\begin{center}
\noindent\includegraphics[width=\textwidth]{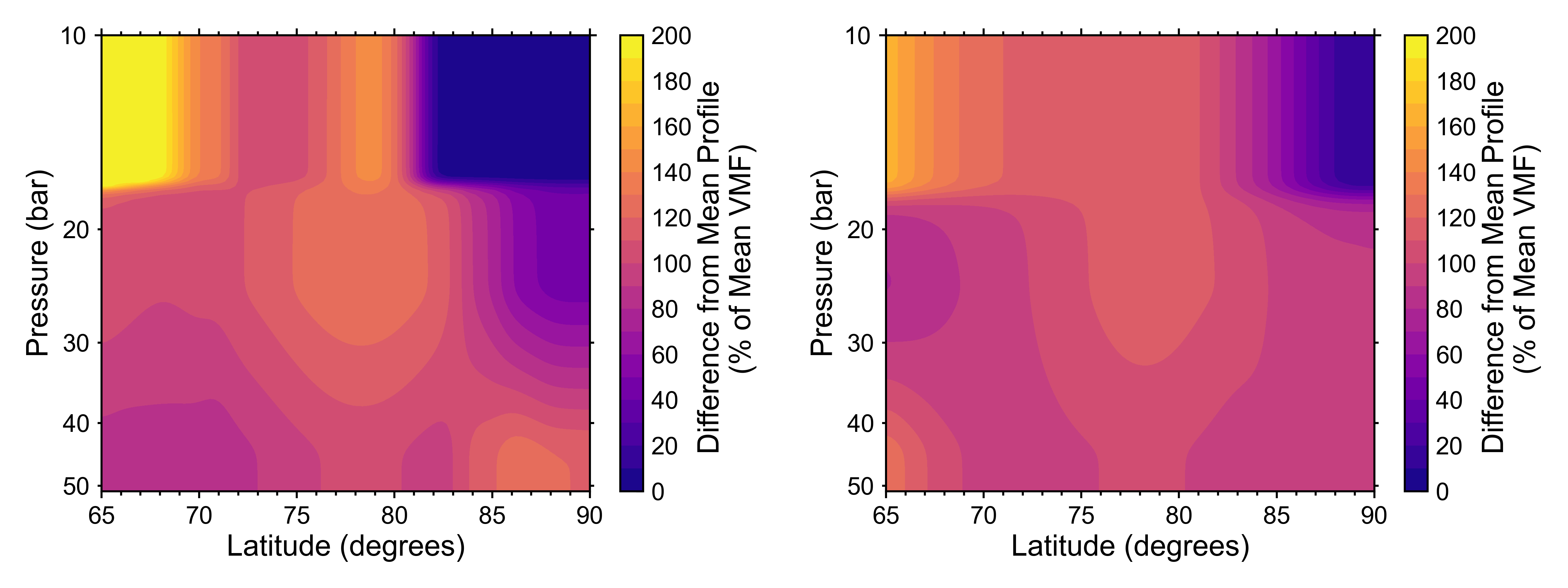}
\caption{Deviation of the retrieved polar abundances (volume mole fractions) as a percentage of the mean (averaged over all latitudes) retrieved profile for (left) H$_2$S and (right) NH$_3$ gas corresponding to end-model 2 (see text). For both charts, the spatial resolution of the inferred quantities range between 4-8 degrees of latitude, and the NH$_4$SH condensation level occurs near 18 bar.}
\label{fig:comp_analysis}
\end{center}
\end{figure}

 First (end-model 1), we assume that composition is constant over the north polar region and adjust the temperature profile to fit brightness temperature spectra over a range of latitudes ($65^\circ \leq \theta \leq 90^\circ)$. From the fit temperature variations as a function of latitude, the vertical wind shear is computed using the thermal wind equation \cite{Pedlosky1987}. Starting from the assumed zonal wind profile of \citeA{Sromovsky2015} at 1.5 bar, the wind shear is integrated with depth to estimate the zonal wind speed at 50 bar. The temperature deviation from the reference profile, vertical wind shear, and zonal wind profiles are shown in Figure \ref{fig:temp_analysis}. Poleward of 70$^\circ$, the temperature decreases below 20 bar to match the dark collar, and at 85$^\circ$, the temperature increases sharply to match the polar spot. The inferred zonal wind around the polar spot increases with depth while the zonal wind speed towards the center of the spot does not increase significantly, consistent with a cyclonic circulation structure. Next (end-model 2), we assume that the temperature is constant and adjust the profiles of either H$_2$S or NH$_3$ gas independently to match the latitude-dependent spectra, with the retrieved distributions shown in Figure \ref{fig:comp_analysis}. Since the volume mole fractions of these gases change by multiple orders of magnitude over this pressure range, these distributions are displayed as percent difference from the mean profile averaged over all latitudes. Gas abundances increase below 20 bar to match the spectrum of the dark collar and are significantly depleted over the polar spot above 20 bars. Uncertainties in these estimates for horizontal variability in both temperature and composition are provided in Supporting Information.
  
 \section{Discussion} 
The bright spot and dark collar at the north pole of Uranus were detected on several days at different wavelengths with consistency independent of imaging approach (e.g. using different visibility weightings, starting models, self-calibration solutions, etc.). Their consistent detection provides strong evidence that these features are real and not imaging artifacts. Both of these features exhibit wavelength dependence, with the central polar spot contrast increasing at shorter wavelengths and the collar contrast increasing at longer wavelengths. Additionally, these observation suggest (albeit at the 2$\sigma$ level) that the polar spot has brightened since 2015, which would imply a strengthening of the polar cyclone as Uranus approaches solstice.  While these images are zonally smeared, unsmearing techniques \cite{Sault2004} are not attempted, since the longitude-resolved maps of 2015 observations presented by \citeA{Molter2021} do not exhibit significant variation in longitude. Our ECCM fits to the lower latitude spectra suggest reasonable agreements with the recent results of \citeA{Molter2021}, and we expect further analysis of Uranus observations over a longer record to provide insights in seasonal variability in a more general sense. ECCM fits to spectra at polar latitudes, however, disagree with the results of \citeA{Molter2021} in requiring an H$_2$S-dominant solution. This is not entirely surprising, as these observations more densely sample the wavelength range of Uranus' spectrum which achieve the poorest fit in \citeA{Molter2021} Figure 9. While the ECCM constraints can match lower latitude spectra well, they appear less able to adequately fit the polar spectrum, and future work should consider in greater detail how different model assumptions impact the ability to fit polar observations specifically. 

The overall accuracy in the brightness temperature spectra determined from the pipeline-calibrated VLA observations is set by the accuracy of the absolute flux calibrator (order 3-5\%), and this determines the statistical uncertainty in the retrieved vertical profiles presented in Figure \ref{fig:vert_prof}. Uncertainties in the spatial brightness temperature distribution within each image, however, are determined by the image sensitivities. In all cases, the off-source image noise is an order of magnitude lower than the uncertainty in the absolute brightness temperature. This motivates our focus in Figures \ref{fig:temp_analysis} and \ref{fig:comp_analysis} on the changes with latitude in retrieved temperature and composition, which can in principle be determined with greater accuracy than the vertical structure. For all retrievals, the polar spot was chosen as the reference location (i.e. non-retrieved quantities were set to the values retrieved over the polar spot using the MCMC/ECCM fits) since it is the brightest region on the planet and requires the lowest absorber abundances. Fitting scaled composition profiles at lower polar latitudes then requires only the addition of either H$_2$S or NH$_3$, whereas the opposite approach would require the abundances of both absorbers to be subtracted to fit brighter regions. Additionally, our approach assumes that the absolute magnitude of the retrieved temperature and composition does not have a large effect on the inference of variability with latitude from the observations. This is likely accurate when ECCM constraints are imposed but will be less accurate if the atmospheric structure deviates significantly from these models. The finite angular resolution of the observations (shown in the bottom left corner of the images in Figure \ref{fig:uranus_images}) will artificially smooth the distribution of retrieved quantities with latitude and therefore impact the determination of their variability.

The derived variations in temperature as a function of latitude (end-model 1, Figure \ref{fig:temp_analysis}) suggest a warmer temperature within the polar spot above the NH$_4$SH cloud level and a slight temperature depression below this cloud in the colder polar collar. The assumption of geostrophic balance allows us to use the gradients in temperature with latitude to derive profiles of vertical shear in the zonal wind, and we determine an estimate of the 50 bar zonal wind profile in the limiting case that temperature drives all variations in polar brightness temperature. We note that the derived zonal wind profile, while qualitatively appropriate within the necessary assumption of geostrophy, may suffer from biases due to the necessity to integrate wind shear down from a starting altitude of 1.5 bar and the fact that our observations lose sensitivity to temperature and opacity variations above 5 bar generally. If changes in atmospheric temperature are the dominant factor in elevated microwave brightness temperatures observed at the polar spot (i.e. end-model 1 is closer to the truth), then the thermal wind calculation implies zonal wind speed increases westwards with depth at the periphery (i.e. within 5 degrees) of the polar spot, consistent with the presence of a polar cyclone.  Assuming that compositional gradients (end-model 2, Figure \ref{fig:comp_analysis}) drive the brightness temperature structure, changes with latitude in either NH$_3$ or H$_2$S abundances are consistent in the requirement of an increase in opacity between 75-80$^\circ$ and a decrease in opacity over the polar spot. We note that while the results in Figures \ref{fig:temp_analysis} and \ref{fig:comp_analysis} suggest that the NH$_4$SH cloud base serves a boundary between different regimes of horizontal variability, the precise location of this boundary is not quantitatively constrained here and is related to the retrieval grid choice (see Supporting Information). Generally, fits to spectra appear consistent with a boundary region within approximately $\pm$ 5 bar of the predicted NH$_4$SH cloud layer. While prior explanations for the equator-pole contrast in microwave emission observed on Uranus postulate a mean descending meridional circulation over the polar cap, the variability in composition with latitude illustrated by Figure \ref{fig:comp_analysis} suggests that smaller-scale cells may exist that locally enhance gas abundances at 80$^\circ$. Since H$_2$S is less microwave-opaque per molecule than NH$_3$, the percentage variability shown in Figure \ref{fig:comp_analysis} implies larger variations in mole fraction across latitude for this gas (mean profiles are similar to those determined via MCMC fitting at 70$^\circ$ and shown in Supporting Information). In Supporting Information, we present estimates of uncertainty in Figures \ref{fig:temp_analysis} and \ref{fig:comp_analysis} and the minimum achieved cost functions. In all cases, fits at the polar spot are poorer than those at lower polar latitudes, which further motivates our suggestion of a more detailed assessment of atmospheric models applicable to Uranus' polar thermal emission.   

 Elevated temperatures, stronger peripheral zonal circulation, and a drier atmosphere are all hallmarks of cyclonic vortex circulation, as previously observed on Saturn and Neptune \cite{Fletcher2018, Fletcher2014, DePater2014, Tollefson2021a}. If the north polar spot is indeed a cyclonic vortex, then a combination of physical temperature and compositional differences is likely to cause the elevated brightness temperatures.  While the vertical motion at the poles is likely net descending, our observations suggest that small-scale meridional cells might be present at depth. The connection of these brightness temperature structures with patterns observed in images at other wavelengths is unclear and should be the subject of future work. ALMA observations presented by \citeA{Molter2021} do not exhibit the emission trough and central spot observed in the VLA observations, suggesting local confinement of these features deeper than the 10 bar level. The central polar spot in our observations appears to have the same horizontal extent as the slightly off-center polar spot identified by \citeA{Karkoschka2015} from analysis of Voyager ISS observations, and this spot was found to rotate two hours faster than regions immediately surrounding it. However, mid-IR observations sensitive to comparable pressure ranges (0.5 - 1 bar) have not observed a corresponding change in emission relative to other polar latitudes that would indicate the presence of a compact cyclone \cite{Flasar1987, Roman2020}. While the location of the dark collar in the VLA observations is also close to a low reflectivity feature in Keck 1.6 $\mu$m observations of Uranus' south pole, the relationship between these features is ambiguous, since the same feature observed by the VLA in 2015 is not present in 2015 Keck observations \cite{Sromovsky2012, Sromovsky2019}. Interestingly, the 13 $\mu$m thermal maps presented by \citeA{Roman2020}, which are sensitive to both stratospheric polar thermal polar emission and acetylene abundance, exhibit greater morphological similarity to our Figure 1 than their 18 $\mu$m observations, which are sensitive to tropospheric temperature (although the locations of peaks and troughs in polar emission are offset by several degrees). The diversity of Uranus' morphology across wavelengths is generally suggestive of a multi-layered meridional circulation structure \cite{Fletcher2020}, and further efforts should focus on synthesizing recent results to derive further insight into Uranus' atmospheric processes.  

\section{Conclusions}
In this letter, we present new 0.7 cm - 5 cm VLA observations of Uranus resulting in detailed views of Uranus' polar structure. These observations confirm the detection of a dark collar at 75$^\circ$-80$^\circ$ latitude observed in 2015 data described by \citeA{Molter2021} (and marginally by \citeA{Hofstadter1990}), and find that the central spot is significantly brighter than the rest of the pole in several wavelength bands. We find that while the bright spot contrast decays with depth, the dark collar contrast increases with depth. We considered the range of variations in atmospheric thermal structure and composition that could reproduce this feature on Uranus. In the case that all polar brightness temperature variations are the result of changes in temperature, we also determine estimates of zonal winds that suggest cyclonic circulation at the polar spot, whereas the alternate case (brightness temperature variations are caused only by changes in composition) suggests gas depletion at least above the NH$_4$SH cloud. It is plausible that both processes are occurring by analogy with Cassini CIRS observations of Saturn's polar cyclone \cite{Fletcher2018} and observations of Neptune's polar vortex \cite{Fletcher2014, Roman2022}. At lower polar latitudes, inferences of temperature and composition gradients both suggest that small-scale circulation cells are superimposed on the net descending motion characteristic of polar meridional circulation, enriching the dark collar in gas. Additionally, comparison of 2 cm observations between 2015 and 2022 suggest a possible strengthening of the polar cyclone in recent years. As Uranus continues towards summer solstice, we encourage multi-wavelength observations from the microwave to the visible to monitor the evolution of Uranus' polar atmospheric state so that further insight can be gained in the study of giant planet circulation processes. Further 2D or 3D modeling of Uranus' polar circulation is also encouraged to further constrain the formation of such a feature. Additionally, the inclusion of a microwave radiometer on future missions to Uranus could result in significant improvements to the results reported here via reduction of uncertainties in absolute calibration and improvements in achievable spatial resolution at longer wavelengths.

\section{Open Research}
%AGU requires an Availability Statement for the underlying data needed to understand, evaluate, and build upon the reported research at the time of peer review and publication.

%Authors should include an Availability Statement for the software that has a significant impact on the research. Details and templates are in the Availability Statement section of the Data and Software for Authors Guidance: \url{https://www.agu.org/Publish-with-AGU/Publish/Author-Resources/Data-and-Software-for-Authors#availability}

%It is important to cite individual datasets in this section and, and they must be included in your bibliography. Please use the type field in your bibtex file to specify the type of data cited. Some options include Dataset, Software, Collection, ComputationalNotebook. Ex: 
%\\
%\begin{verbatim}
%
%@misc{https://doi.org/10.7283/633e-1497,
%  doi = {10.7283/633E-1497},
%  url = {https://www.unavco.org/data/doi/10.7283/633E-1497},
%  author = {de Zeeuw-van Dalfsen, Elske and Sleeman, Reinoud},
%  title = {KNMI Dutch Antilles GPS Network - SAB1-St_Johns_Saba_NA P.S.},
%  publisher = {UNAVCO, Inc.},
%  year = {2019},
%  type = {dataset}
%}

% \end{verbatim}

% For physical samples, use the IGSN persistent identifier, see the International Geo Sample Numbers section:
% \url{https://www.agu.org/Publish-with-AGU/Publish/Author-Resources/Data-and-Software-for-Authors#IGSN}
%%%%%%%%%%%%%%%%%%%%%%%%%%%%%%%%%%%%%%%%%%%%%%%
Raw VLA data used to generate images of Uranus from this report are available from the NRAO Data Archive at \url{data.nrao.edu} under project 21B-301. Data reduction and imaging were performed using the CASA software package (version 6.5.2), which is available at \url{https://casa.nrao.edu/}. Additionally, all processed images shown in Figure 1 are available at https://doi.org/10.5281/zenodo.7690111

\acknowledgments
We acknowledge comments from two anonymous reviewers which improved the manuscript. This work was carried out at the Jet Propulsion Laboratory, California Institute of Technology, under contract with NASA and sponsored by the JPL Research and Technology Development Fund. This paper makes use of VLA data obtained under project VLA/21B-301. The National Radio Astronomy Observatory (NRAO) is a facility of NSF operated under cooperative agreement by Associated Universities, Inc. IdP and NM are supported by NASA grant NNX16AK14G through the Solar System Observations (SSO) program to the University of California, Berkeley.

%% ------------------------------------------------------------------------ %%
%% References and Citations

%%%%%%%%%%%%%%%%%%%%%%%%%%%%%%%%%%%%%%%%%%%%%%%
%
% \bibliography{<name of your .bib file>} don't specify the file extension
%
% don't specify bibliographystyle

% In the References section, cite the data/software described in the Availability Statement (this includes primary and processed data used for your research). For details on data/software citation as well as examples, see the Data & Software Citation section of the Data & Software for Authors guidance
% https://www.agu.org/Publish-with-AGU/Publish/Author-Resources/Data-and-Software-for-Authors#citation

%%%%%%%%%%%%%%%%%%%%%%%%%%%%%%%%%%%%%%%%%%%%%%%

%\bibliography{enter your bibtex bibliography filename here}

\bibliography{library.bib}

%Reference citation instructions and examples:
%
% Please use ONLY \cite and \citeA for reference citations.
% \cite for parenthetical references
% ...as shown in recent studies (Simpson et al., 2019)
% \citeA for in-text citations
% ...Simpson et al. (2019) have shown...
%
%
%...as shown by \citeA{jskilby}.
%...as shown by \citeA{lewin76}, \citeA{carson86}, \citeA{bartoldy02}, and \citeA{rinaldi03}.
%...has been shown \cite{jskilbye}.
%...has been shown \cite{lewin76,carson86,bartoldy02,rinaldi03}.
%... \cite <i.e.>[]{lewin76,carson86,bartoldy02,rinaldi03}.
%...has been shown by \cite <e.g.,>[and others]{lewin76}.
%
% apacite uses < > for prenotes and [ ] for postnotes
% DO NOT use other cite commands (e.g., \citet, \citep, \citeyear, \citealp, etc.).
% \nocite is okay to use to add references from your Supporting Information
%

\end{document}